\documentclass{aipproc}

\layoutstyle{8x11double}

\title{Localization of GRBs by Bayesian Analysis of Data from the HETE WXM}

\author{Carlo Graziani}
{
address={Department of Astronomy \& Astrophysics, University of Chicago,
5640 South Ellis Avenue, Chicago, IL 60637},
email={carlo@oddjob.uchicago..edu},
}

\author{Donald Q. Lamb}
{
address={Department of Astronomy \& Astrophysics, University of Chicago,
5640 South Ellis Avenue, Chicago, IL 60637},
email={lamb@oddjob.uchicago..edu},
}

\author{The HETE Science Team}
  {address={
    An international collaboration of institutions including 
  MIT, LANL, U. Chicago, U.C. Berkeley, U.C. Santa Cruz (USA),
  CESR, CNES, Sup'Aero (France), RIKEN, NASDA (Japan), TESRE (Italy),
  INPE (Brazil), TIFR (India) 
  },
}

\begin{abstract}
We describe a new method of transient point source localization for
coded-aperture X-ray detectors that we have applied to data from the HETE
Wide-Field X-Ray Monitor (WXM).  The method is based upon the calculation
of the likelihood function and its interpretation as a probability density
for the transient source location by an application of Bayes' Theorem. 
The method gives a point estimate of the source location by finding the
maximum of this probability density, and credible regions for the source
location by choosing suitable contours of constant probability density. 
We describe the application of this method to data from the WXM, and give
examples of GRB localizations which illustrate the results that can be
obtained using this method.
\end{abstract}

\begin{document}

\maketitle

\section{Introduction}

The HETE Wide-Field X-Ray Monitor (WXM) is composed of two crossed,
one-dimensional coded-aperture cameras.  GRB locations are inferred from
position histograms in the X and Y cameras, using the known mask pattern
and detector response.

Several processing methods for coded-aperture data have been proposed,
including cross correlation \citep{fenimore78}, least-squares fitting
\citep{doty88}, and Maximum Entropy \citep{sims80,willingdale84}. 
Skinner and Nottingham \citep{skinner93} have described a
maximum-likelihood fitting technique.

In \citep{graziani97} we presented a Bayesian scheme for analyzing
coded-aperture data from such transient events.  The method is based on
the calculation of the joint likelihood function for two stretches of
data:  the stretch covering the transient event itself, and a stretch
before and/or afterwards, which provides information about the
background.  We interpret the likelihood thus obtained as a posterior
probability density for the transient event location by an application of
Bayes' Theorem.

We have implemented the method in the HETE data-processing pipeline, where
it is routinely used to obtain GRB locations.  Here we present the
implementation details, and some of the results that have been obtained to
date.

\section{Implementation}

When high-resolution WXM data from a HETE trigger are received on the
ground, background and burst time intervals are automatically determined
by SNR maximization.  The software then essentially compares the
background-to-burst change in the WXM position histogram data with a Monte
Carlo point source model, using the method described in \citep{graziani97}
to calculate the posterior probability density at each interrogated source
location.

Computation of the Monte Carlo point source model (or ``template'') is
somewhat time-consuming, so that it is not possible to pre-compute
templates in a dense grid spanning all possible locations in the
field-of-view.  We have therefore adopted a two-stage approach:

\begin{itemize}

\item {\sl Coarse Location} --- coarsely spaced pre-computed templates are
compared to the data to produce a location accurate to about 1/3$^\circ$.

\item {\sl Fine Location} --- a freshly-generated template at the coarse
location is shifted around the immediate neighborhood of that location, to
map the posterior probability distribution.

\end{itemize}

\begin{figure}[t]
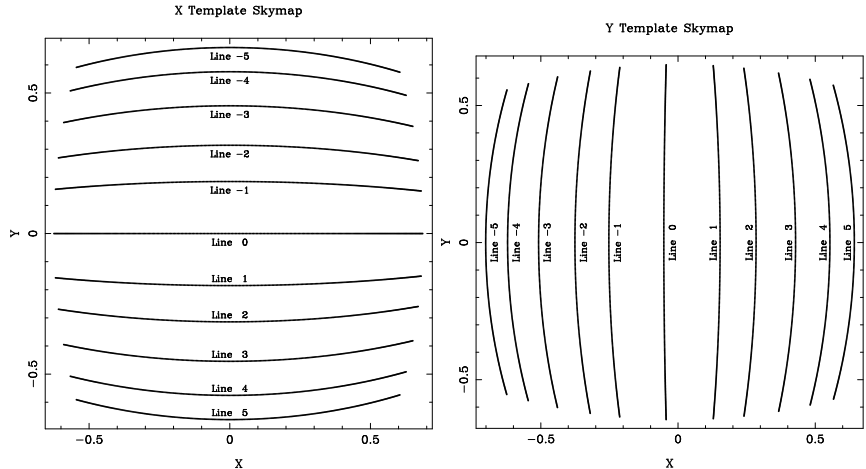


\resizebox{0.7\textwidth}{!}{
\includegraphics[scale=0.3]{skypos_x.ps}
\includegraphics[scale=0.3]{skypos_y.ps}
}
\caption{Lines of Templates for coarse location.  The plots are equal-area
projections, with an opening half-angle of about 40 degrees.}
\label{bftf}

\end{figure}

\subsection{Coarse Location}

The WXM is composed of two 1-dimensional coded-aperture cameras that
resolve locations in perpendicular directions (X and Y).  We can use
1-dimensional arrays of sky locations with pre-computed templates in order
to get coarse X and Y locations.  Given such an array, the code simply
computes the posterior density at each location and plots it as a function
of template number, reporting the highest value.

\begin{figure}[ht]
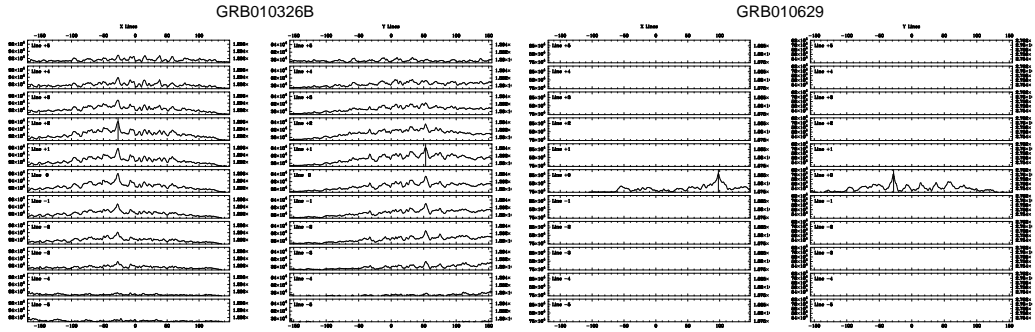


\resizebox{0.8\textwidth}{!}
{
\includegraphics[angle=-90]{llplot_GRB010326b_2up.ps}
\hspace{1.0in}
\includegraphics[angle=-90]{llplot_GRB010629_2up.ps}
}
\caption{Coarse location log-likelihood profiles.  Each set of two panels
shows X and Y localizations.  Left: GRB010326B, 22-line array of
templates.  Right: GRB010629, ``Cross'' configuration of templates.}
\label{coarse}

\end{figure}

One such array that we occasionally use has 22 lines of templates --- 11
each in X and in Y.  The lines are chosen so as to expose select
combinations of adjacent WXM detector wires (the detector in each WXM
camera has six wires, some of which might not be illuminated by a point
source at a certain location).  This allows the X-detector to supply some
Y information, and vice-versa.  Figure \ref{bftf} shows a skymap of this
template configuration.

Another template array is a simple cross - two lines of sky locations, one
in the X direction and one in the Y direction, intersecting at the center
of the FOV.  This array uses less information, but can be run much faster,
and is usually adequate for the purpose of obtaining a coarse localization.

In both cases the spacing between adjacent templates on a line is about
18', which is the fundamental resolution element of WXM.  This is thus
also the accuracy of a coarse location obtained in this fashion.

Figure \ref{coarse} shows examples of the output from the coarse
location procedure, for GRB010326B and GRB010629. Each set of two panels
shows the log-likelihood profile as a function of template number, for the
X and Y template lines.  The two panels on the left show the result of the
analysis of GRB010326B performed using the full 22-line template array.
The two panels on the right show the result of the analysis of GRB010629
performed using the cross template array.  In both cases the best-fit X
and Y angles are marked by a vertical line.

\subsection{Fine Location}

The posterior density function allows us to go beyond merely obtaining a
point estimate of the location --- we can also use it to infer {\sl
contours} of prescribed probability content around the best-fit location. 
That is, we can use it to get error boxes.

\begin{figure}[ht]

\includegraphics[scale=0.45,bb=100 50 510 720]{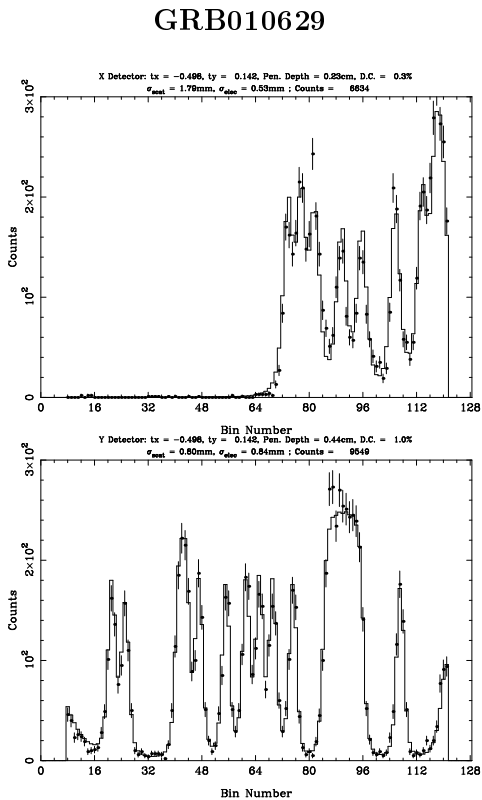}
\caption{Fit of empirical model to Monte Carlo simulation.  The assumed
direction of the source is the coarse location of GRB010629.}
\label{fit}
\end{figure}

We proceed by calculating one very bright MonteCarlo template at the
best-fit coarse location.  We fit that template with an empirical model
constructed by smearing and transforming the coded mask pattern.  This
transformation is designed to give a simplified account of physical
processes such as scattering and detector penetration.  An example of such
a fit is shown in Figure \ref{fit}, which was produced in the analysis of
GRB010629.

We then calculate the posterior density on a $20\times 20$ grid of points
in the neighborhood of the coarse location, using at each point a template
calculated by smearing the mask pattern with the best-fit empirical
parameters determined by the fit of the model to the MonteCarlo template.

\begin{figure}[ht]

\includegraphics[scale=0.3]{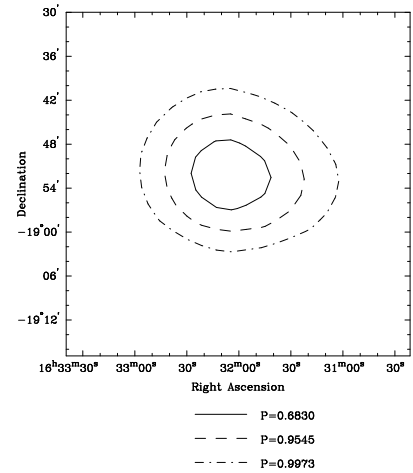}
\caption{Skymap showing location probability contours obtained by the fine
location of GRB010629.}
\label{chart}
\end{figure}

We issue not only a refined best-fit location, but also constant-density
contours containing 68.3\%, 95.5\%, and 99.7\% (statistical) probability
of including the correct location. Figure \ref{chart} shows the location
contours produced in this way for GRB010629.

For the purposes of quick communication, and to simplify follow-up
observations, we also produce ``circularized'' error-boxes by calculating
the radius of a circles about the best-fit location that just contain
68.3\%, 95.5\% and 99.7\% (statistical) probability of including the
correct location.  Since these are not constant probability contours, they
necessarily subtend slightly larger solid angles than the iso-density
contours.  The difference is usually not very large.

The statistical probabilities inferred here must be supplemented by the
systematic location error deduced from our calibration studies.

\end{document}